\documentclass[12pt,preprint]{aastex}
\begin{document}

\title{\bf Comprehensive Observations of a Solar Minimum CME with STEREO}

\author{B. E. Wood\altaffilmark{1}, R. A. Howard\altaffilmark{1},
  S. P. Plunkett\altaffilmark{1}, D. G. Socker\altaffilmark{1}}

\altaffiltext{1}{Naval Research Laboratory, Space Sciences Division,
  Washington, DC 20375; brian.wood@nrl.navy.mil}

\begin{abstract}

     We perform the first kinematic analysis of a CME observed
by both imaging and in~situ instruments on board STEREO,
namely the SECCHI, PLASTIC, and IMPACT experiments.
Launched on 2008 February 4, the CME is tracked
continuously from initiation to 1 AU using the SECCHI imagers
on both STEREO spacecraft, and is then detected by the PLASTIC and
IMPACT particle and field detectors on board STEREO-B.  The CME is
also detected in~situ by ACE and SOHO/CELIAS at Earth's L1
Lagrangian point.  The CME hits STEREO-B, ACE, and SOHO on
2008 February 7, but misses STEREO-A entirely.  This event provides
a good example of just how different the same event can look when viewed from
different perspectives.  We also demonstrate many ways in which
the comprehensive and continuous coverage of this CME by STEREO
improves confidence in our assessment of its kinematic behavior, with
potential ramifications for space weather forecasting.
The observations provide several lines of evidence in favor of the
observable part of the CME being narrow in angular extent, a
determination crucial for deciding how best to convert observed
CME elongation angles from Sun-center to actual Sun-center distances.

\end{abstract}


\section{INTRODUCTION}

     The {\em Solar Terrestrial Relations Observatory} (STEREO) mission
is designed to improve our understanding of coronal mass ejections (CMEs)
and their interplanetary counterparts ICMEs (``interplanetary CMEs'') in
many different ways.  Consisting of two spacecraft observing the Sun from
very different locations, STEREO simultaneously observes the Sun and
interplanetary medium (IPM) from two vantage points, allowing a much
better assessment of a CME's true three-dimensional structure from the
two-dimensional images.  STEREO has the capability of
observing CMEs far into the IPM thanks to its two Heliospheric Imagers,
HI1 and HI2, which can track CMEs all the way to 1~AU.  The only
other instrument with comparable capabilities is the Solar Mass
Ejection Imager (SMEI) on the {\em Coriolis} spacecraft, which is still
in operation \citep{cje03,bvj04,dfw06,tah08}.  Finally, the
two spacecraft possess particle and field instruments that can study
ICME properties in~situ.  The ability to continuously follow a CME from
the Sun into the IPM actually blurs the distinction between the CME and
ICME terms.  Since most of this paper will be focused on white-light
images of the CME, we will generally use only the CME acronym.

     \citet{raha08} reported on the first CMEs observed by
STEREO that could be continuously tracked into the IPM by HI1 and HI2.
They tracked the events over $40^{\circ}$ from the Sun.  We extend
this work further by presenting observations of a CME that can be tracked
all the way to 1~AU, where the event is then detected by particle and
field detectors on one of the two spacecraft.  It is also detected by
the {\em Advanced Composition Explorer} (ACE), and by the Charge, Element,
and Isotope Analysis System (CELIAS) on board the {\em Solar and
Heliospheric Observatory} (SOHO).  Both ACE and SOHO are at Earth's L1
Lagrangian point, so the CME's detection there means that it qualifies
as an Earth-directed event.  This CME is
therefore useful for illustrating how STEREO's unique perspective can
provide a much better assessment of the kinematics and structure of
potentially geoeffective CMEs.  This will become more important as we
move away from the 2008 solar minimum and strong Earth-directed CMEs
become more frequent.

\section{THE STEREO INSTRUMENTS}

     The two STEREO spacecraft were launched on 2006~October~26,
one into an orbit slightly inside that of Earth (STEREO-A), which
means that it moves ahead of the Earth in its orbit,
and one into an orbit slightly outside that of Earth (STEREO-B), which
means that it trails behind the Earth.  Since launch the
separation of the A and B spacecraft has been gradually growing.
Figure~1 shows their locations on 2008~February~4,
which is the initiation date of the CME of interest here.  At this
point STEREO A and B had achieved a separation angle of $45.3^{\circ}$
relative to the Sun.

     The two STEREO spacecraft contain identical sets of instruments.
The imaging instruments are contained in a package called the
Sun-Earth Connection Coronal and Heliospheric Investigation (SECCHI),
which will be described in more detail below.  There are two
in~situ instruments on board, the Plasma and Suprathermal Ion
Composition (PLASTIC) instrument \citep{abg08}, and the
In-situ Measurements of Particles and CMEs Transients (IMPACT)
package \citep{mha08,jgl08}.  The former
studies the properties of ions in the bulk solar wind, and the latter
studies electrons, energetic particles, and magnetic fields in
the IPM.  Finally, there is a radio wave detector aboard each
spacecraft called STEREO/WAVES, or SWAVES \citep{jlb08},
but SWAVES did not see any activity relevant to our particular CME.

     Most of the data presented in this paper will be from the five
telescopes that constitute SECCHI, which are fully described by
\citet{raho08}.  Moving from the Sun outwards, these
consist firstly of an Extreme Ultraviolet Imager (EUVI), which
observes the Sun in several extreme ultraviolet bandpasses.  There are
then two coronagraphs, COR1 and COR2, which observe the white light
corona at elongation angles from the Sun of $0.37^{\circ}-1.07^{\circ}$
and $0.7^{\circ}-4.2^{\circ}$, respectively.  These angles correspond
to distances in the plane of the sky of $1.4-4.0$ R$_{\odot}$
for COR1 and $2.5-15.6$ R$_{\odot}$ for COR2.  Finally, there are
the two Heliospheric Imagers, HI1 and HI2, mentioned in \S1, which
observe the white light IPM in between the Sun and Earth at
elongation angles from the Sun of $3.9^{\circ}-24.1^{\circ}$ and
$19^{\circ}-89^{\circ}$, respectively.
At these large angles, plane-of-sky distances become very misleading,
so we do not quote any here.
Figure~1 shows explicitly the overlapping COR2, HI1,
and HI2 fields of view for STEREO-A and STEREO-B on 2008~February~4.

\section{THE 2008 FEBRUARY 4 CME}

\subsection{Imaging the Event}

     Figures~2--6 provide examples of images of the February~4 event
from all five of the SECCHI telescopes.  Figure~2 shows sequences
of images in two of the four bandpasses monitored by EUVI:  the He~II
$\lambda$304 bandpass and the Fe~XII $\lambda$195 bandpass.
Note that the actual time cadence is 10 minutes in both of these
bandpasses, rather than the 30 minute time separation of the chosen
images in Figure~2.

     At about 8:16 UT a prominence is observed to be gradually
expanding off the southeast limb of the Sun in the EUVI-A $\lambda$304
images.  This expansion then accelerates into a full prominence eruption,
as a small flare begins at about 8:36 in the He~II and Fe~XII images
some distance northwest of the prominence.  The flaring site is
indicated by an arrow in Figure~2.  This is not a strong flare
in EUVI, and there is no GOES X-ray event recorded at all at this time,
so the flare is apparently too weak to produce sufficient high
temperature plasma to yield a GOES detection.

     Figures~3 and 4 show white-light COR1 and COR2 images of a
CME that emerges shortly after the EUVI flare begins.  As was the
case for EUVI, the actual time cadence is three times faster than
implied by the selected images:  10 minutes for COR1 and 30
minutes for COR2.  The synoptic COR1 and COR2 programs actually involve
the acquisition of 3 separate images at three
different polarization angles, which we combine into a single 
total-brightness image for our purposes.  [Technically, the COR2
cadence is actually 15 minutes, alternating between the acquisition
of full polarization triplets, which we use here, and total brightness
images computed from polarization doublets combined onboard,
which we do not use \citep{raho08}.]  The coronagraph images
are all displayed in running-difference mode in Figures~3 and 4, where
the previous image is subtracted from each image.  This
is a simple way to subtract static coronal structures and emphasize
the dynamic CME material.

     From the perspective of STEREO-A the CME is first seen by COR1-A
off the southeast limb, as expected based on the location of the flare
and prominence eruption.  However, the strong southern component to
the CME motion seen in the COR1-A images in Figure~3 disappears by the
time the CME leaves the COR2-A field of view.  In the final COR2-A image
in Figure~4 the CME is roughly symmetric about the ecliptic plane,
in contrast to its COR1-A appearance.
\citet{hc04} have noted that near solar minimum, CMEs appear to be
deflected towards the ecliptic plane, presumably due to the presence of
high speed wind and open magnetic field lines emanating from polar
coronal holes.  The February~4 CME may be another example of this.

     The CME's appearance is radically different from the point of view
of STEREO-B, illustrating the value of the multiple-viewpoint STEREO
mission concept.  The EUVI-B flare is only $10^{\circ}$ from disk center,
so the expectation is that any CME observed by STEREO-B will be a halo
event directed at the spacecraft.  However, the COR1-B and COR2-B
images show only a rather faint front expanding slowly in a southwesterly
direction, though the COR2-B movies do provide hints of expansion at
other position angles, meaning that this might qualify as a partial
halo CME.  It is possible that if the CME was much brighter it might have
been a full halo event.  It is impressive how much fainter the event is
from STEREO-B than from STEREO-A, possibly due to the CME subtending
a larger solid angle from STEREO-B's perspective, with some of the CME
blocked by the occulter.
The visibility of a CME as a function of viewing angle can also
be affected by various Thomson scattering effects \citep{mda02,av06}.

     Both COR1-A and COR1-B (see Fig.~3) show a CME directed more to the
south than would be expected based on the flare site (see Fig.~2), and
COR1-B shows more of a westward direction than would be expected
considering how close to disk center the flare is from the point of
view of STEREO-B.  We speculate that perhaps the coronal hole just east
of the flare site (see EUVI-B Fe~XII $\lambda$195 images in Fig.~2)
plays a role in deflecting the CME into the more southwesterly trajectory
seen by STEREO-B.  Thus, this CME seems to show evidence for two
separate deflections from coronal holes:  the initial deflection to
the southwest from the low latitude hole adjacent to the flare site,
and the more gradual deflection back towards the ecliptic plane seen
in COR2-A (see Fig.~4).

     Figures~5 and 6 show HI1 and HI2 images of the CME as it
propagates through the IPM to 1~AU.  As was the case for COR1 and
COR2, the HI1 and  HI2 data are displayed in running difference mode.
The time cadence of HI1 and HI2 data acquisition are 40 minutes and
2 hours, respectively.

     The large fields of view of the HI telescopes
and the increasing faintness of CME fronts as they move further from
the Sun make subtraction of the stellar background a very important
issue.  For HI1 we first subtract an average image computed from about
2 days worth of data encompassing the Feb.~4 event.  This removes
the static F-corona emission, which eliminates the large brightness
gradient in the raw HI1 images.  We then use a simple median filtering
technique to subtract the stars before the running difference subtraction
of the previous image is made.  Artifacts from some of the brightest
stars are still discernible in Figure~5, including vertical streaks due
to exposure during the readout of the detector.
Median filtering does not work well for the diffuse
background produced by the Milky Way, so the Milky Way's presence
on the right side of the HI1-B images is still readily apparent.
A somewhat more complicated procedure is used for HI2, which
involves the shifting of the previous image before it is subtracted to
make the running-difference sequence, in an effort to better eliminate
the stellar background.  This method should be effective for both the
diffuse Milky Way and stellar point source background, but median
filtering is also used to further improve the stellar subtraction.
The HI2 image processing procedure is described in more detail
by \citet{nrs08b}.

     The bright CME front is readily apparent in the HI1-A images, but in
HI1-B the CME can only be clearly discerned in the lower left corner of the
last two images in Figure~5.  This is consistent with expectations from
the appearance of the CME in the COR2 data.  The situation becomes more
complicated in the HI2 field of view (FOV).  Figure~6 shows two HI2-A images,
and also shows the positions of Earth, SOHO, and STEREO-B in the FOV.
Earth and SOHO are behind a trapezoidal occulter, which is used
to prevent the image from being contaminated by a very overexposed
image of Earth.  The first image shows that the CME front is initially a
bright, semicircular front, consistent with its appearance in HI1-A.  But
it quickly fades, becoming much harder to follow.  There are other fronts
in the FOV (see Fig.~6) associated with a corotating
interaction region (CIR), which confuses things further.  The CME front
appears to overtake the CIR material and the second image in Figure~6
shows the CME as it approaches the position of STEREO-B.
At this point the front is much more well defined in the southern
hemisphere than in the north.

     Given the potential confusion between our CME front and the
CIR material, it is worthwhile to briefly review what CIRs are
and how they are perceived by STEREO.  \citet{nrs08a,nrs08b} have
already described CIR fronts seen by HI2 in some detail, which
have been the most prominent structures regularly seen by HI2 in STEREO's
first year of operation.  The CIRs are basically standing waves of
compressed solar wind, where high speed wind coming from low latitude
coronal holes is running into low speed wind.  The CIRs stretch outwards
from the Sun in a spiral shape due to the solar rotation, and have
a substantial density enhancement that HI2-A sees as a
gradually outward propagating front (or series of fronts) as the CIR
rotates into view.  The HI2-B imager does {\em not} see the approach
of the CIR in the distance like HI2-A does because HI2-B is looking at
the west side of the Sun rotating away from the spacecraft instead of the
east side rotating towards it, where HI2-A is looking (see
Fig.~1).  Instead, when the CIR reaches STEREO-B, HI2-B seen a very
broad front pass very rapidly through the foreground of the FOV
as the CIR passes over and past the spacecraft.

     Since our CME front appears to overtake a CIR in the HI2-A
images, it is tempting to look for evidence of interaction between
the two.  However, we believe that the leading edge of the CME
is actually always ahead of the CIR.  The appearance of ``overtaking''
is due to a projection effect, where the faster moving CME
is in the foreground while the apparently slower CIR material seen in
Figure~6 is in the background.  Support for this interpretation is
provided by Figure~2.  The EUVI-B Fe~XII $\lambda$195 images show
the coronal hole that is the probable source of the high speed wind
responsible for the CIR.  The coronal hole is just east of the flare region
that represents the CME initiation site, so with respect to the Sun's
westward rotation the CME leads the high speed wind that yields the CIR.
This is the same coronal hole that we suppose to
have deflected the CME into a more southwesterly direction, but the
leading edge of the CME is always ahead of the CIR.  Nevertheless, it is
quite possible that the sides and trailing parts of the CME may be
interacting with the CIR structure.  Trying to find clear evidence for
this in the HI2 data ideally requires guidance from models of CME/CIR
interaction.  Such an investigation is clearly a worthwhile endeavor, but
it is outside the scope of our purely empirical analysis here.

     Returning to the CME, just as HI2-B does not see CIRs until they
engulf STEREO-B, HI2-B does not perceive the February~4 CME
until it is practically on top of the spacecraft (as seen from STEREO-A).
As the CME approaches and passes over STEREO-B, HI2-B sees a very broad,
faint front pass rapidly through the foreground of the FOV,
similar in appearance to the CIR fronts described above.  Though the
rapid front is apparent in HI2-B movies, its faintness combined with
its very broad and diffuse nature makes it practically impossible to
discern in still images, so we have not attempted to show it in any
HI2-B images here.

\subsection{In Situ Observations}

     Figures~2-6 demonstrate STEREO's ability to track a CME continuously
from its origin all the way out to 1~AU using the SECCHI telescopes,
even for a modest event like the February~4 CME.
Figure~7 demonstrates STEREO's ability to study the properties
of the CME when it gets to 1~AU.  The upper two panels of Figure~7
show the solar wind proton density and velocity sampled by the PLASTIC
experiments on both STEREO spacecraft from February~5-17, and the
bottom panel shows the magnetic field strength observed by
IMPACT.  For comparison, we have also added measurements made at
Earth's L1 Lagrangian point by ACE and SOHO/CELIAS.  Including both ACE
and SOHO/CELIAS data provides us with two independent measurements
at L1.  (The CELIAS instrument does not provide magnetic field
measurements, though.)  The CME is detected by STEREO-B, and more weakly
by ACE and SOHO, but it is not detected at all by STEREO-A.

     Given that the CME's initiation site is near Sun-center as seen
by STEREO-B (see Fig.~2), it is not surprising that the CME eventually
hits that spacecraft.  STEREO-B sees a density and magnetic field
increase on February~7 at the same time that HI2-A sees the CME
front reach STEREO-B, so there is good reason to believe that this
is the expected ICME corresponding to the February~4 CME.  However,
the particle and field response are not characteristic of a typical
ICME or magnetic cloud \citep[see, e.g.,][]{lj06}, and it is
difficult to tell exactly where the ICME begins and ends.
The wind velocity increases from an ambient slow solar wind speed of
about 360 km~s$^{-1}$ to the CME's propagation speed of 450 km~s$^{-1}$,
but the velocity increase trails the density and magnetic field increase by
at least 12 hours.  Perhaps much of the field and density
excess associated with the CME may be slow solar wind that has been
overtaken and piled up in front of the original CME front, but we
cannot rule out the possibility that the CME may be mixed up with
some other magnetic structure, confusing the ICME signature in
Figure~7.  Another possibility is that the central axis of the
CME passed to the south of the spacecraft, leading to a more muddled
magnetic field signature.

     The ICME is also detected by ACE and CELIAS.  The velocity profiles
seen at L1 are practically identical to that seen by PLASTIC-B.  The
density profiles seen by ACE and CELIAS are somewhat discrepant.  The
ACE data show a weak, narrow density spike at the time of maximum
density at STEREO-B, while the CELIAS data only show a broad, weak
density enhancement.  In any case, the density enhancement at L1 is
weaker than at STEREO-B.  The magnetic field enhancement seen by ACE is
also weaker than at STEREO-B, and shorter in duration.  Thus,
STEREO-B receives a more direct hit from the CME than ACE and SOHO, which
are $23.6^{\circ}$ away from STEREO-B (see Fig.~1).  This is once again
consistent with the CME's direction inferred from the SECCHI images.
At an angle from STEREO-B of $45.3^{\circ}$, the PLASTIC and IMPACT
instruments on STEREO-A do not see the CME at all, providing a hard
upper limit for the angular extent of the CME in the ecliptic plane.

     It is worthwhile to compare and contrast how HI2 and the in~situ
instruments perceive the CME.  The CME front seen by HI2-A (see Fig.~6)
appears to reach the location of STEREO-B at about 18:00 UT on February~7.
This corresponds roughly to the time when the densest part of
the CME is passing by STEREO-B (see Fig.~7).  However, both the density
and magnetic field data indicate that less dense parts of the CME front
reach STEREO-B much earlier.  This demonstrates that much of the CME
structure is unseen by HI2-A.  The HI2-A front displayed in Figure~6 is
only the densest part of the CME.  For ACE and SOHO there is an even
greater disconnect between the HI2-A CME front and the in~situ
observations of it.  Movies of the fading CME front allow it to just
barely be tracked out to the position of ACE and SOHO, which it reaches
at about 6:00 UT on February~8, well after the weak density enhancement
seen by these instruments is over.  This means that HI2-A does not
see the part of the CME that hits the spacecraft at L1,
only seeing the denser parts of the structure that are farther away than
L1, in the general direction of STEREO-B.

     This leads to the schematic picture of the CME geometry shown in
Figure~1.  Based on the velocity curves in Figure~7, there is essentially
no velocity difference along the CME front, and no time delay between the
CME arrival at STEREO-B and L1, so the
CME front is presumably roughly spherical as it approaches 1~AU, as
shown in Figure~1.  However, we have argued above that HI2-A only sees
the densest part of the CME, which hits STEREO-B, and HI2-A does not see
the foreground part that hits L1 and Earth at all.  The dotted
purple line in Figure~1 crudely estimates the full extent of the CME,
which we know hits STEREO-B, ACE, and SOHO but not STEREO-A, while the
shorter solid line arc is an estimate of the part of the CME that HI2-A
actually sees.  It is difficult to know how far the CME
extends to the right of STEREO-B in Figure~1.  There
is little if any emission apparent to the east of the Sun in COR2-B
movies (see Fig.~4), which is why we have not extended the CME arc
very far to the right of STEREO-B in Figure~1.  Thus, the final
picture is that of a CME that has a total angular extent of no more
than $60^{\circ}$, with the visible part of the CME constituting
less than half of that total.

     Besides showing the CME signatures observed by PLASTIC, IMPACT,
ACE, and CELIAS, Figure~7 also shows these instruments' observations of
the CIR that follows, the presence of which is also apparent in the
HI2-A images in Figure~6, as noted in \S3.1.  All these spacecraft see a
strong density and magnetic field enhancement, which is accompanied by
a big jump in wind velocity as the spacecraft passes from the slow
solar wind in front of the CIR to the high speed wind that trails
it.  Such signatures are typical of CIRs seen by in~situ instruments
\citep[e.g.,][]{nrs08a,nrs08b}.  There is a significant time delay between
when the CIR hits STEREO-B, then ACE and SOHO, and finally STEREO-A.  This
time delay illustrates the rotating nature of the CIR structure.  It is
curious that the time delay is significantly longer between ACE/SOHO and
STEREO-A than it is between STEREO-B and ACE/SOHO despite the angular
separation being about the same (see Fig.~1).  It is also interesting
that the density, velocity, and magnetic field profiles seen by the
three spacecraft are rather different.  Though outside the scope of this
paper, a more in-depth analysis of this and other STEREO-observed CIRs
is certainly worthwhile, especially given the substantial number of
these structures observed by STEREO in the past year.

\subsection{Kinematic Analysis}

     Possibly the simplest scientifically useful measurements one can
make from a sequence of CME images are measurements of the velocity of
the CME as a function of time.  However, even these seemingly simple
measurements are complicated by uncertainties in how to translate
apparent 2-dimensional motion into actual 3-dimensional velocity.
We now do a kinematic analysis of the February~4 CME, and in the
process we show how comprehensive STEREO observations can
improve confidence in such an analysis.

     In order to measure the velocity and acceleration of a CME's
leading edge, positional measurements must first be made from the
SECCHI images.  What we actually measure is not distance but an
elongation angle, $\epsilon$, from Sun center.  Many previous authors
have discussed methods of inferring distance from Sun-center, $r$,
from $\epsilon$ \citep{swk07,tah07,tah08,nrs08b}.  One approach, sometimes
referred to as the ``Point-P Method,'' assumes the CME leading edge is
an intrinsically very broad, uniform, spherical front centered on the
Sun, in which case \citep{tah07}
\begin{equation}
r=d \sin \epsilon.
\end{equation}
Here $d$ is the distance of the observer to the Sun, which is close to
1~AU for the STEREO spacecraft, but not exactly (see Fig.~1).

     Another approach, which \citet{swk07} call the
``Fixed-$\phi$ Method,'' assumes that the CME is a relatively narrow,
compact structure traveling on a fixed, radial trajectory at an angle,
$\phi$, relative to the observer's line of sight to the Sun, in which
case
\begin{equation}
r=\frac{d\sin \epsilon}{\sin(\epsilon+\phi)}.
\end{equation}
[Note that this is a more compact version of equation (A2)
in \citet{swk07}.]  In the top panel of Figure~8, we show plots of
$r$ versus time, as seen from STEREO-A, using both equations (1) and (2).
In the latter case we have assumed the CME trajectory is radial from
the flare site, meaning $\phi=46^{\circ}$.  There is a time gap in the
HI2 measurements, corresponding to when the CME front is too confused
with CIR material to make a reliable measurement.

     The bottom panel of Figure~8 shows velocities computed from
the distance measurements in the top panel.  Velocities
computed strictly from adjacent distance data points often lead to
velocities with huge error bars, which vary wildly in time
in a very misleading fashion.  For this reason, as we compute
velocities from the distances point-by-point we actually
skip distance points until the uncertainty in the computed velocity
ends up under some assumed threshold value (70 km~s$^{-1}$ in this case),
similar to what we have done in past analyses of SOHO data \citep{bew99}.
The velocity uncertainties are computed assuming the following
estimates for the uncertainties in the distance measurements:
1\% fractional errors for the COR1 and COR2 distances,
and 2\% and 3\% uncertainties for HI1 and HI2, respectively.

     There are significant differences in the distance and velocity
measurements that result from the use of equations (1) and (2).
In order to explore the reasons behind the distance differences,
first note that a point in an image represents a direction
vector in 3D space.  If this vector has a closest approach to the Sun at
some point P, the geometry assumed by the Point-P method always
assumes that this point P represents the real 3D location of the
apparent leading edge seen by the observer.  Thus, distances estimated
using equation (1) by definition represent a lower bound on
the actual distance \citep{tah07}, explaining why the Point-P
data points are always at or below the Fixed-$\phi$ data
points in Figure~8.

     The two methods lead to different inferences about the CME's
kinematic behavior.  The Fixed-$\phi$ method implies an acceleration
up to a maximum velocity of about 700 km~s$^{-1}$ in the COR2 FOV,
followed by a gradual deceleration through HI1 and into HI2.  In
contrast, the Point-P method suggests that the CME accelerates to
about 500 km~s$^{-1}$ in the COR2 FOV and then continues to accelerate
more gradually through HI1 and into HI2, before decelerating
precipitously in HI2.  However, this last precipitous deceleration
is clearly an erroneous artifact of the Point-P geometry, which
assumes that the CME has a very broad angular extent, encompassing
all potentially observed position angles relative to the Sun, and
implicitly assuming that the CME engulfs the observer when it reaches
1~AU.  That is why equation (1) does not even allow the possibility
of measuring $r$ greater than 1~AU.  However, we know that the
Feb.~4 CME does {\em not} hit the observer (i.e., STEREO-A).

     We have argued near the end of \S3.2 that the Feb.~4 CME does
not have a very large angular extent, and that the extent of the
observed part of the CME is even more limited (see Fig.~1).  Thus,
the Fixed-$\phi$ geometry is a much better approximation
for this particular event.  It is important to note that this
conclusion will not be the case for broader, brighter CMEs, where
the Point P approach might work better.  The Fixed-$\phi$ method does
have the disadvantage that it requires a reasonably accurate knowledge
of $\phi$, though the known flare location provides a good estimate.
And as the CME travels outwards, there will still be some degree of
uncertainty introduced by the likelihood that the observed leading
edge is not precisely following precisely the same part of the CME
structure at all times.

     The effects of these uncertainties can be explored by comparing
the CME velocities measured in the HI2-A FOV with the in~situ velocity
observed by PLASTIC-B.  If the uncertainties are low, the SECCHI
image-derived velocities should agree well with the PLASTIC-B velocity.
The Fixed-$\phi$ velocities measured in the HI2 FOV in Figure~8 (at
times of $t\gtrsim30$~hr) average around 530 km~s$^{-1}$,
somewhat higher than the 450 km~s$^{-1}$ velocity seen by PLASTIC-B
(see Fig.~7).  This is presumably indicative of the aforementioned
systematic uncertainties.  Figure~9 illustrates how the discrepancy can
be addressed by lowering the assumed CME trajectory angle, $\phi$.
Figure~9 plots $r$ versus $\phi$ for many values of $\phi$, computed using
equation (2).  The curves steepen in the HI2 FOV
($\epsilon=19^{\circ}-89^{\circ}$) as $\phi$ increases, meaning that
velocities inferred from these distances will also increase.  Thus,
lowering $\phi$ below the $\phi=46^{\circ}$ value assumed in
Figure~8 will lower the inferred HI2 velocities.

     In order to determine which $\phi$ value works best, we perform a
somewhat more sophisticated kinematic analysis than that in Figure~8.
Compared to the point-by-point analysis used in Figure~8, a cleaner and
smoother velocity profile can be derived from the data if the distance
measurements are fitted with some functional form, which in essence
assumes that the timescale of velocity variation is long compared to
the time difference between adjacent distance measurements.
Polynomial or spline fits are examples of such functional forms that
can used for these purposes.  However, we ultimately decide on a
different approach, relying on a very simple physical model of the
CME's motion.  This model assumes an initial acceleration for the CME,
$a_1$, which persists until a time, $t_1$, followed by a second
acceleration (or deceleration), $a_2$, lasting until time $t_2$,
followed finally by constant velocity.  This model also has two
additional free parameters:  a starting height, and a time shift of
the model distance-time profile to match the data.  The two-phase
model bears some resemblance to the ``main'' and ``residual''
acceleration phases of a CME argued for by \citet{jz06}.  But to us
the appeal of this simple model is that not only are its parameters
physical ones of interest, it also seems to fit the data as well or
better than more complex functional forms, despite having only six
free parameters.

     Figure~10 shows our best fit to the data using this model.
The top panel shows the leading edge distances computed assuming
$\phi=38^{\circ}$, which turns out to be the value that leads to
the observed PLASTIC-B velocity of 450 km~s$^{-1}$ in the HI2-A FOV.
The solid line shows our best fit to the data, determined using a
chi-squared minimization routine, where we have assumed the
same fractional errors in the distance measurements as we did
in Figure~8 (see above).  With these assumed uncertainties, the best
fit ends up with a reduced chi-squared of $\chi_{\nu}^2=1.33$.
This agrees well with the $\chi_{\nu}^2\approx 1$ value expected
for a good fit \citep{prb92}, which implies that the error bars
assumed for our measurements are neither unrealistically small nor
unreasonably large.

     The bottom two panels of the figure show the velocity and
acceleration profiles implied by this fit.  The velocity
at 1~AU (214~$R_{\odot}$) in the HI2-A FOV ends up at 450 km~s$^{-1}$
as promised.  It should be emphasized that in forcing the HI2-A velocity
to be consistent with the PLASTIC-B measurement, we are implicitly assuming
that the part of the CME front being observed by HI2-A has the same
velocity as the part of the CME front that hits STEREO-B.  Essentially,
this amounts to assuming that the CME front is roughly spherical and
centered on the Sun at 1~AU, as pictured in Figure~1 and argued for
in \S3.2.  The excellent agreement between the CME velocity seen by
PLASTIC-B and that seen at L1 by ACE and SOHO/CELIAS also implies that
this assumption is a very good one for this event.  But this may not
be the case for all events, so comparing HI2-A and PLASTIC-B velocities
may not always be appropriate.

     The $\phi=38^{\circ}$ value assumed in Figure~10 is $8^{\circ}$
less than the $\phi=46^{\circ}$ value that radial outflow from the
observed flare site would suggest.  This result could indicate that
the CME's overall center-of-mass trajectory is truly at least $8^{\circ}$
closer to the STEREO-A direction than the flare site would predict.
(More if there is a component of deflection perpendicular to the
ecliptic plane.)  In \S3.1 we noted that the COR1-B and COR2-B images
imply a deflection of the CME into a more southwesterly trajectory than
suggested by the flare site, possibly due to the adjacent coronal hole.
The western component of this deflection would indeed predict a CME
trajectory less than $\phi = 46^{\circ}$ angle suggested by the flare.
Thus, interpreting the $8^{\circ}$ shift as due to this deflection is
quite plausible.

     However, this interpretation comes with two major caveats.
One is that the part of the CME seen as the leading edge by HI2-A
is not necessarily representative of either the geometric center of the
CME, or its center-of-mass.  Measurements
from a location different from that of STEREO-A could in principle see
a different part of the CME front as being the leading edge, thereby
leading to a different trajectory measurement.  The second caveat is
the aforementioned issue of the observed leading edge not necessarily
faithfully following the same part of the CME front at all times,
which could yield velocity measurement errors and therefore an
erroneous $\phi$ measurement.

     Figure~10 represents our best kinematic model of the February~4
CME, which can be described as follows.  The model suggests that the
CME's leading edge has an initial acceleration of $a_1=159$ m~s$^{-2}$
for its first $t_1=1.1$ hours, reaching a maximum velocity of
689 km~s$^{-1}$ shortly after entering the COR2 FOV.  Until $t_2=33$
hours the CME then gradually decelerates at a rate of
$a_2=-2.1$ m~s$^{-2}$ during its journey through the COR2 and HI1
fields of view, eventually reaching its final coast velocity of
451 km~s$^{-1}$ shortly after reaching the HI2 FOV, this velocity being
consistent with the PLASTIC-B measurement.

     Interaction with the ambient solar wind is presumably responsible
for the $a_2$ deceleration inferred between 0.024 and 0.47~AU, as the
PLASTIC-B data make it clear that the CME is traveling through slower
solar wind plasma.  Note that the Point-P measurements in Figure~8 are
not only inconsistent with this $a_2$ IPM deceleration, but they would
actually imply an {\em acceleration} of the CME at that time.
This emphasizes the importance of the issue of
how to compute distances from elongation angles.  Even basic qualitative
aspects of a CMEs IPM motion, such as whether it accelerates or
decelerates, depend sensitively on this issue.  Given that the CME
is plowing through slower solar wind material, an IPM deceleration
seems far more plausible than an acceleration.  This is yet another argument
that the Fixed-$\phi$ geometry is better for this particular event
than the Point-P geometry.

\subsection{Implications for Space Weather Prediction}

     An event like the February~4 CME is perfect for assessing the
degree to which the unique viewpoint of the STEREO spacecraft can
yield better estimates of arrival times for Earth-directed CMEs.  The
February~4 CME is directed at STEREO-B, so STEREO-B's in~situ
instruments tell us exactly when the CME reaches 1~AU, but the STEREO-B
images of the event close to the Sun provide very poor velocity estimates
by themselves because of the lack of knowledge of the CME's precise
trajectory.  For full halo CMEs, \citet{rs05} provide a prescription to
determine the true expansion velocity from its lateral expansion
\citep[see also][]{rs06}.  But the uncertainties remain large, and in any
case this prescription is not helpful for our February~4 event, which is
barely perceived as a partial halo, let alone a full halo.  Therefore, it
is STEREO-A that by far provides the best assessment of the CME's kinematic
behavior thanks to its location away from the CME's path.  And it is
STEREO-A that is therefore in a much better position to predict
ahead of time when the event should reach STEREO-B.

     To better quantify this, we imagine a situation where only STEREO-B
data is available.  The CME has just taken place and the CME has been
observed by COR1-B and COR2-B as shown in Figures~3 and 4.  We can then
ask the question, what would our estimated CME velocity
be from the STEREO-B data alone and what would be the predicted
arrival time at 1~AU?  The apparent plane-of-sky velocity of the CME
in the COR2-B images is about 240 km~s$^{-1}$ (assuming $\phi=90^{\circ}$).
The total travel time to Earth at this speed is about 174 hours, leading
to a predicted arrival time of roughly Februrary~11, 15:00 UT.  This
is 4 days after the actual arrival time on February~7, so this prediction
is obviously very poor!

     If the EUVI-B flare location is used to provide
an estimated trajectory of $\phi=10^{\circ}$, the velocity
estimated from the COR2-B data increases dramatically to about
1000 km~s$^{-1}$.  In this case the 1~AU travel time decreases
to only 42 hours, corresponding to a predicted arrival time
of February~6, 3:00 UT.  This is well over a day {\em before} the
actual arrival time.  For events directed at the observing
spacecraft, CME velocity measurements are particularly sensitive to
uncertainties in the exact trajectory angle.  There is
also the problem that the observed leading edge motion in the COR2-B
images may have more to do with the lateral expansion of the CME rather
than the motion outwards from the Sun.

     If a similar thought experiment is done for the
STEREO-A data, the COR2-A images alone and an assumption of
the $\phi=46^{\circ}$ trajectory suggested by the EUVI-A flare
site lead to a CME velocity in COR2-A of about 590 km~s$^{-1}$.
This corresponds to a 1~AU travel time of about 71 hours, leading to
a predicted arrival time at 1~AU of February~7, 8:00 UT.  This is only
a few hours after the arrival of the CME suggested by the
IMPACT-B magnetic field data (see Fig.~7), though it is about 13
hours before the peak density seen by PLASTIC-B, which is what the CME
front observed by the SECCHI imagers actually corresponds to
(see \S3.3).  Improving the arrival time prediction of the peak
CME density would require taking into account the deceleration
of the CME during its travel through the IPM (see Fig.~10).

     It is clear that STEREO-A's perspective provides a dramatic
improvement in our ability to predict when the February~4 CME
reaches STEREO-B.  An analysis of multiple events like this
one would allow this improvement to be better quantified.
In the spirit of previous analyses such as \citet{ng01},
perhaps an analysis of multiple events such as this one would also
provide empirical guidance in how to predict the deceleration
during IPM travel (or perhaps acceleration in some cases), which
is clearly necessary to achieve arrival time estimates that are
good to within a few hours.  The ability of SECCHI to provide
continuous tracking information on CMEs could in principle allow
time-of-arrival estimates to be continously improved during a
CME's journey to 1~AU.

\section{SUMMARY}

     We have presented STEREO observations of a CME that occurred in
the depths of the 2008 solar minimum, when there were not many of
these events taking place.  The February~4 CME is not a particularly
dramatic event, but it has an advantageous trajectory.  It is directed
at STEREO-B, so that it eventually hits that spacecraft and is
detected by its in~situ instruments.  A different part of the CME
hits the ACE and SOHO spacecraft at Earth's L1 Lagrangian point.
The CME's trajectory is far enough away from the STEREO-A direction
that STEREO-A images can provide an accurate assessment of the CME's
kinematic behavior, which is not possible from STEREO-B's location.
This event illustrates just how much the appearance of a CME can
differ between the two STEREO spacecraft, which at the time had
an angular separation relative to the Sun of $45.3^{\circ}$, a
separation that continues to increase with time by about $44^{\circ}$
per year.

     Despite the relative faintness of the event, the SECCHI imagers are
still able to track it continuously all the way from the Sun to 1~AU,
which provides hope that as the Sun moves towards solar maximum, STEREO
will be able to provide similarly comprehensive observations of many
more such CMEs.  The kinematic analysis presented here is the first
based on such a comprehensive STEREO data set, involving both SECCHI
images and in~situ data, but hopefully many others will follow.
We have used two different methods of computing CME leading edge
distances from measured elongation angles:  1. The Point-P method,
which assumes the CME is a broad, uniform, spherical front; and
2. The Fixed-$\phi$ method, which assumes a narrow, compact CME
structure traveling radially from the Sun.  Our analysis illustrates
just how sensitive conclusions about the kinematic behavior of a CME
are to the method used.  The first method suggests continued
acceleration in the IPM, while the second implies a deceleration.
Fortunately, the comprehensive nature of observations of the
February~4 CME has provided us with an abundance of evidence that the
observable part of this CME has a very limited angular extent.
Therefore, the Fixed-$\phi$ method is clearly best in this case,
leading to our best kinematic model for the CME in Figure~10.  But we do
not expect the Fixed-$\phi$ method to necessarily be the best
option for all STEREO-observed CMEs.

     Finally, the geometry of the event allows us to use the two
spacecrafts' observations to quantify just how much more accurately
the CME's arrival time at 1~AU can be predicted using images taken
away from the CME's path (from STEREO-A in this case), compared to
images taken from directly within it (from STEREO-B in
this case).  The STEREO-A prediction proves to be dramatically
better than STEREO-B's.  Thus, STEREO could in principle be able to
improve space weather forecasting for Earth-directed events in the
coming years.

\acknowledgments

We would like to thank Neil Sheeley and Peter Schroeder for
helpful discussions and assistance in this project.
The STEREO/SECCHI data are produced by a consortium of NRL (US),
LMSAL (US), NASA/GSFC (US), RAL (UK), UBHAM (UK), MPS (Germany), CSL
(Belgium), IOTA (France), and IAS (France).  In addition to funding
by NASA, NRL also received support from the USAF Space Test Program
and ONR.  In addition to SECCHI, this work has also made use of data
provided by the STEREO IMPACT and PLASTIC teams, supported by NASA
contracts NAS5-00132 and NAS5-00133.  We have also made use of data
provided by the CELIAS/MTOF experiment on SOHO, which is a joint
ESA and NASA mission.  We thank the ACE SWEPAM and MAG instrument
teams and the ACE Science Center for providing the ACE data.

\clearpage

\begin{figure}[p]
\plotone{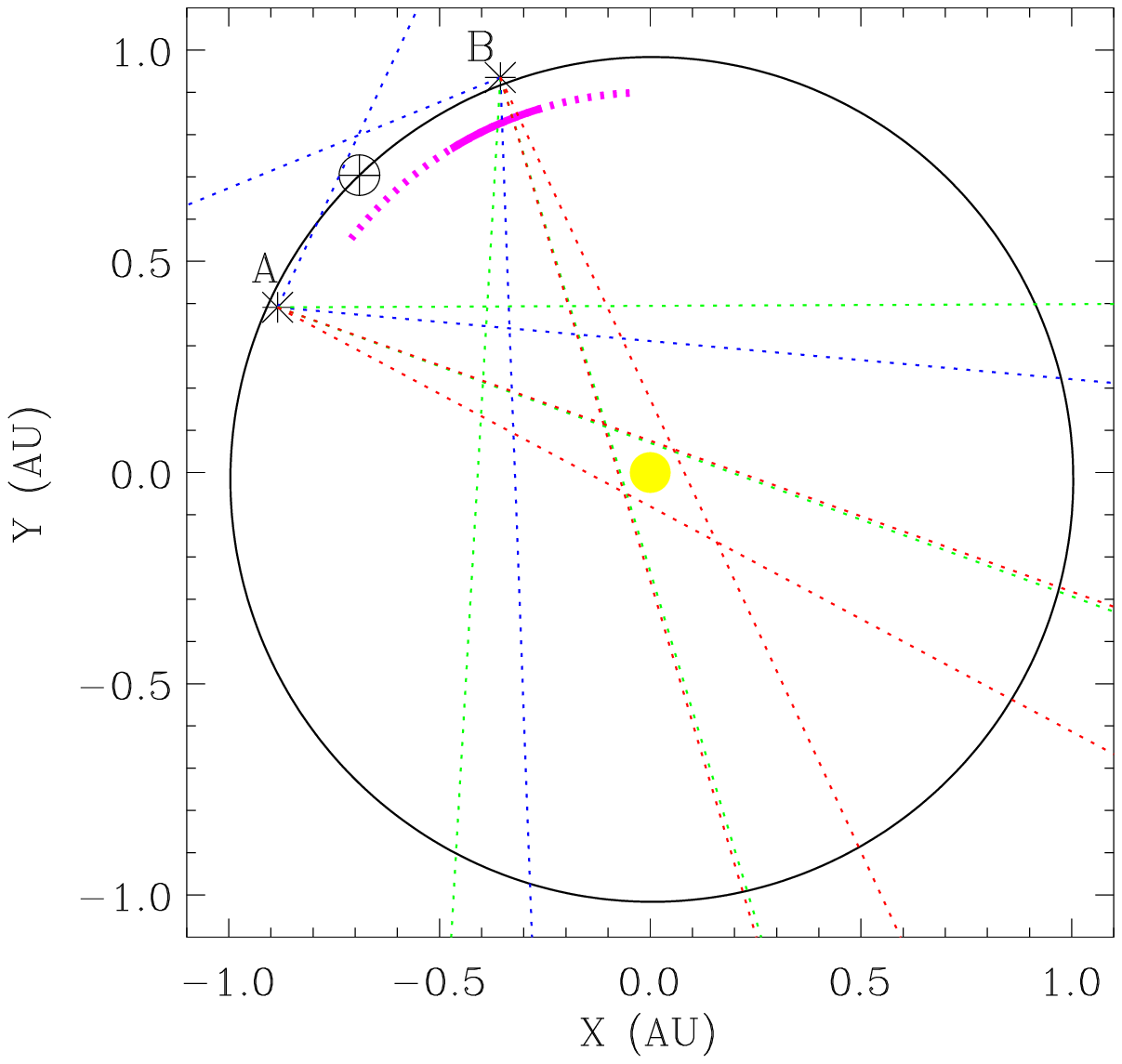}
\caption{The locations of Earth, STEREO-A, STEREO-B, and the Sun
  (at the origin) on 2008~February~4 in heliocentric aries ecliptic
  coordinates.  The red, green, and blue dotted lines indicate the
  fields of view of the COR2, HI1, and HI2 telescopes on board
  STEREO-A and B.  The purple arc is an estimated location for the
  Feb.~4 CME's leading edge as it approaches 1~AU, where the part of
  the arc represented as a solid
  line is the part of the CME that we detect in SECCHI images from
  STEREO-A
  images, and the dotted line indicates the parts of the CME front
  that we do not see (see \S3.2).}
\end{figure}

\begin{figure}[p]
\plotone{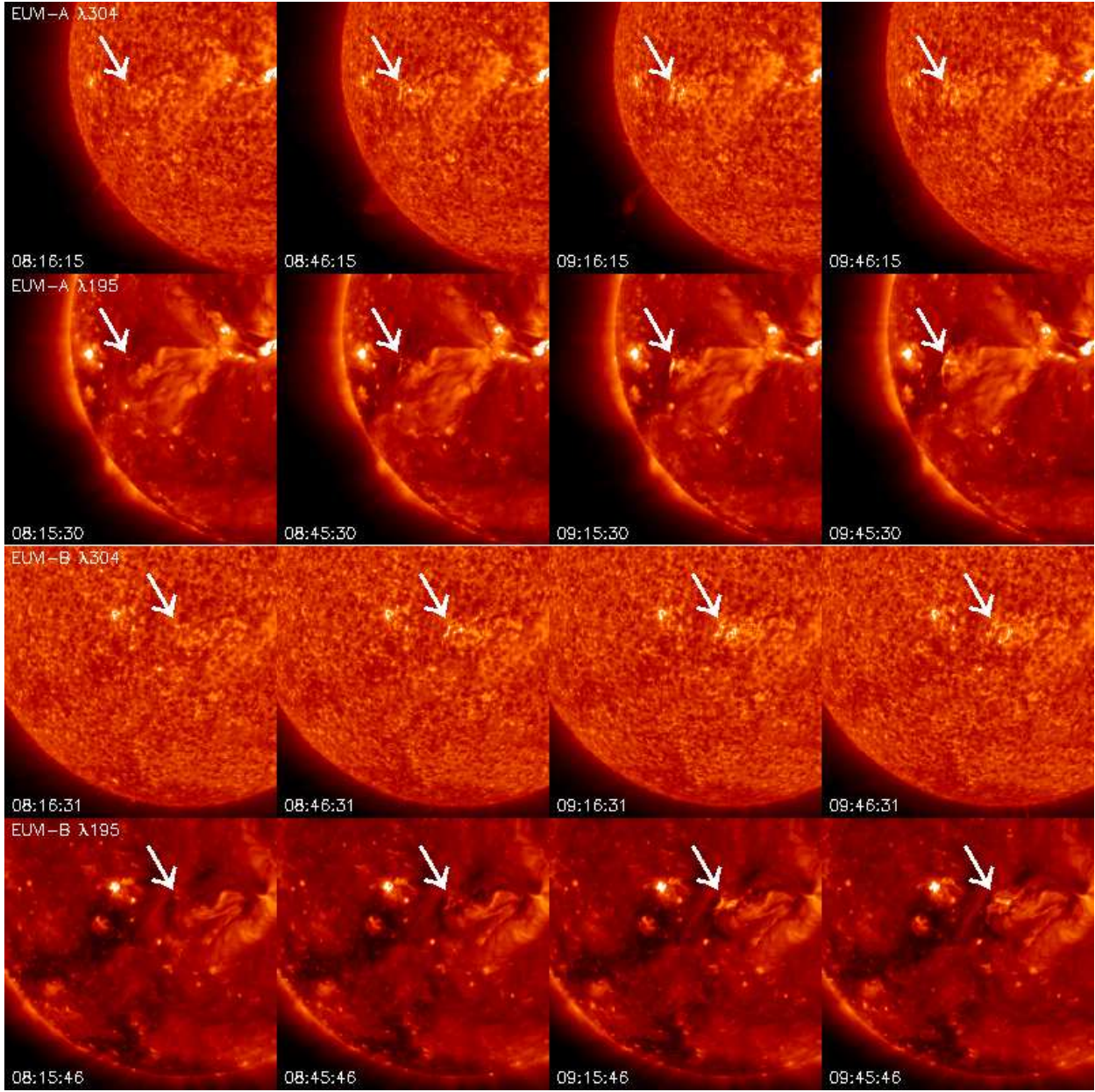}
\caption{Sequences of EUVI images taken near the beginning of the
  2008 Feb.~4 CME.  The upper 2 sequences are He~II $\lambda$304
  and Fe~XII $\lambda$195 images from STEREO-A and the bottom 2
  sequences are He~II and Fe~XII images from STEREO-B.  The arrows
  point to a region that flares weakly during the event.
  The EUVI-A He~II $\lambda$304 images also show a prominence eruption
  off the southeast limb.}
\end{figure}

\begin{figure}[p]
\plotone{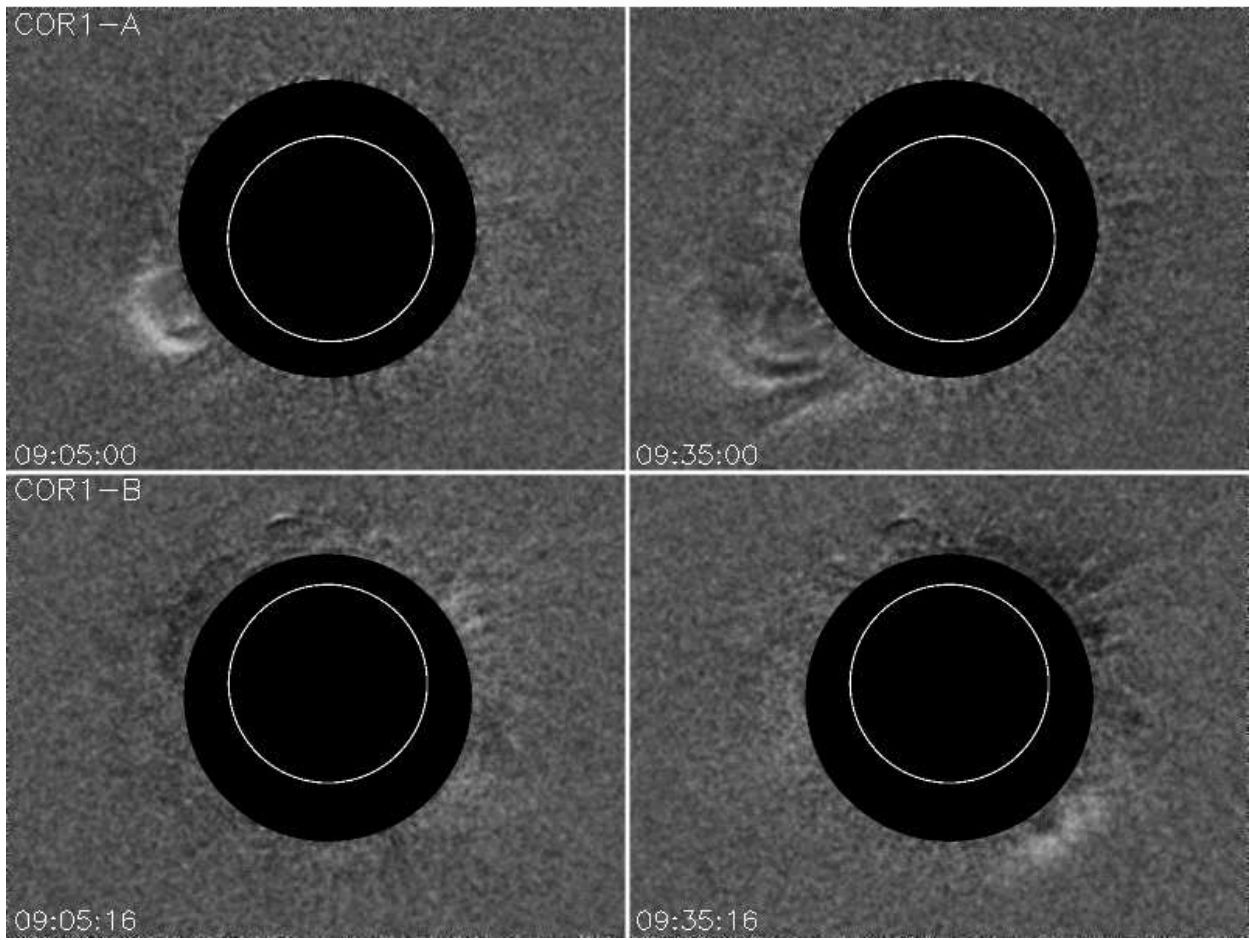}
\caption{Running-difference COR1 images from STEREO-A (top) and STEREO-B
  (bottom) showing the 2008 Feb.~4 CME erupting off the southeast limb in
  COR1-A, but primarily off the southwest limb in COR1-B.  The white circle
  indicates the location of the solar disk.}
\end{figure}

\begin{figure}[p]
\plotone{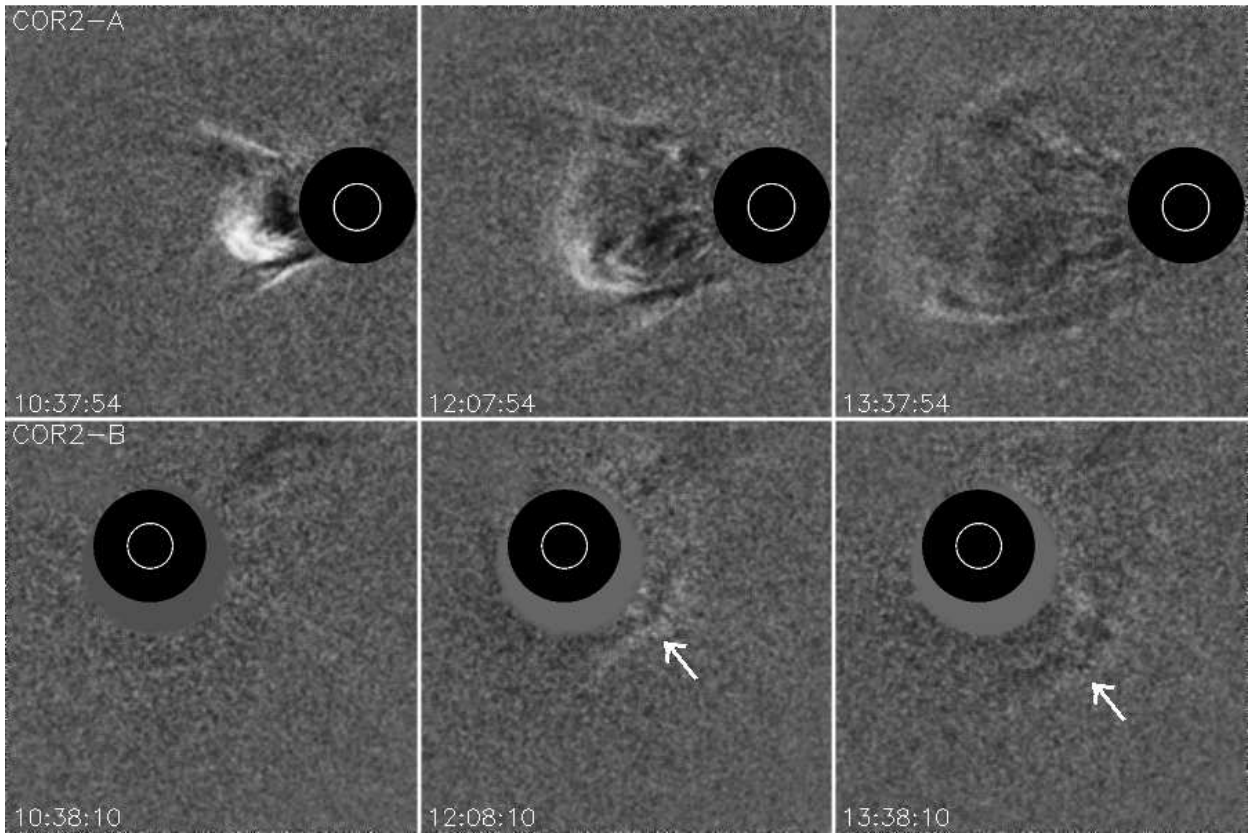}
\caption{Running-difference COR2 images from STEREO-A (top) and STEREO-B
  (bottom) showing the 2008 Feb.~4 CME.  The CME is clearly seen off the
  east limb in COR2-A, but it is much fainter and primarily off the
  southwest limb in COR2-B (arrows).  The white circle
  indicates the location of the solar disk.  Beyond the occulter there
  is some additional masking for COR2-B to hide blooming caused by a
  slight miscentering of the Sun behind the occulting disk.}
\end{figure}

\begin{figure}[p]
\plotone{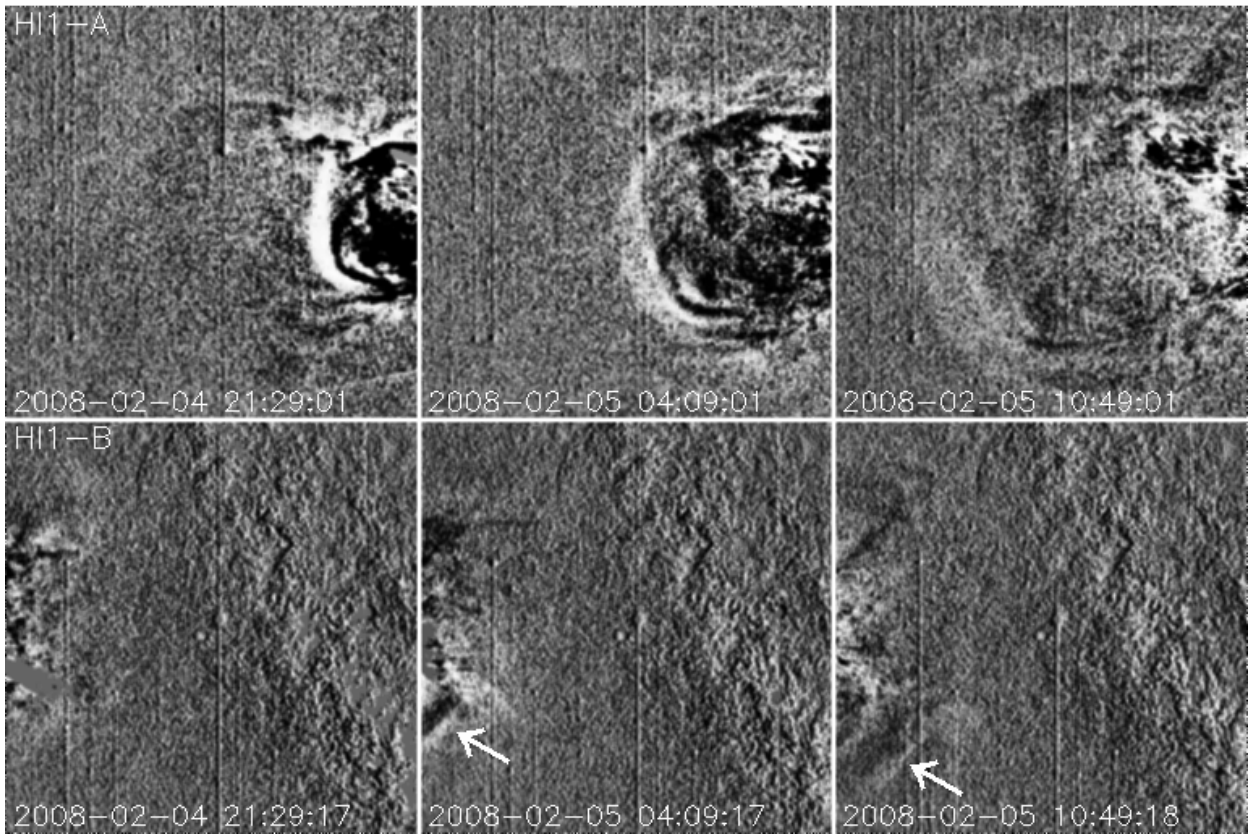}
\caption{Running-difference HI1 images from STEREO-A (top) and STEREO-B
  (bottom) showing the 2008 Feb.~4 CME.  The Sun is to the right in the
  HI1-A images and to the left for HI1-B (see Fig.~1).  The CME front is
  obvious in HI1-A, but is only faintly visible in the lower
  left corner of the last two HI1-B images (arrows).}
\end{figure}
\begin{figure}[p]

\plotone{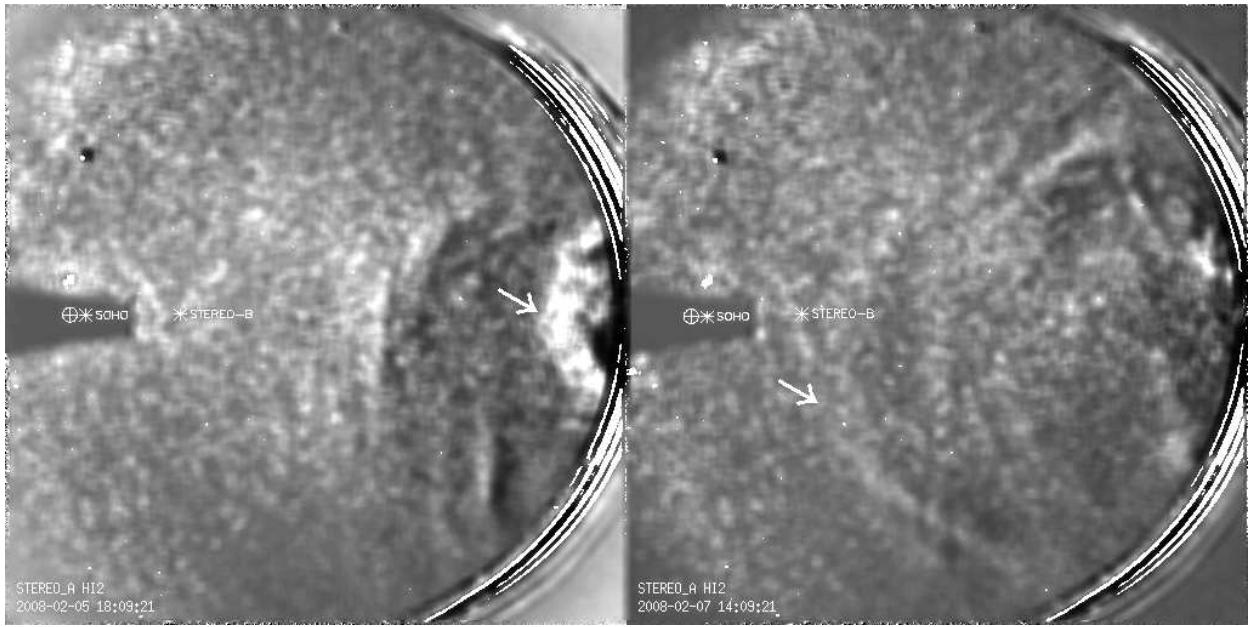}
\caption{Running-difference HI2 images from STEREO-A showing
  the 2008 Feb.~4 CME (arrows).  The positions of the Earth, SOHO,
  and STEREO-B are also shown.  The first image shows the bright CME
  front as it enters the field of view on Feb.~5, but the CME front quickly
  fades and becomes confused with a CIR structure in the background,
  which is gradually rotating towards the observer.  The second
  image shows the CME front just before it crosses the apparent
  position of STEREO-B on Feb.~7.}
\end{figure}

\begin{figure}[p]
\plotone{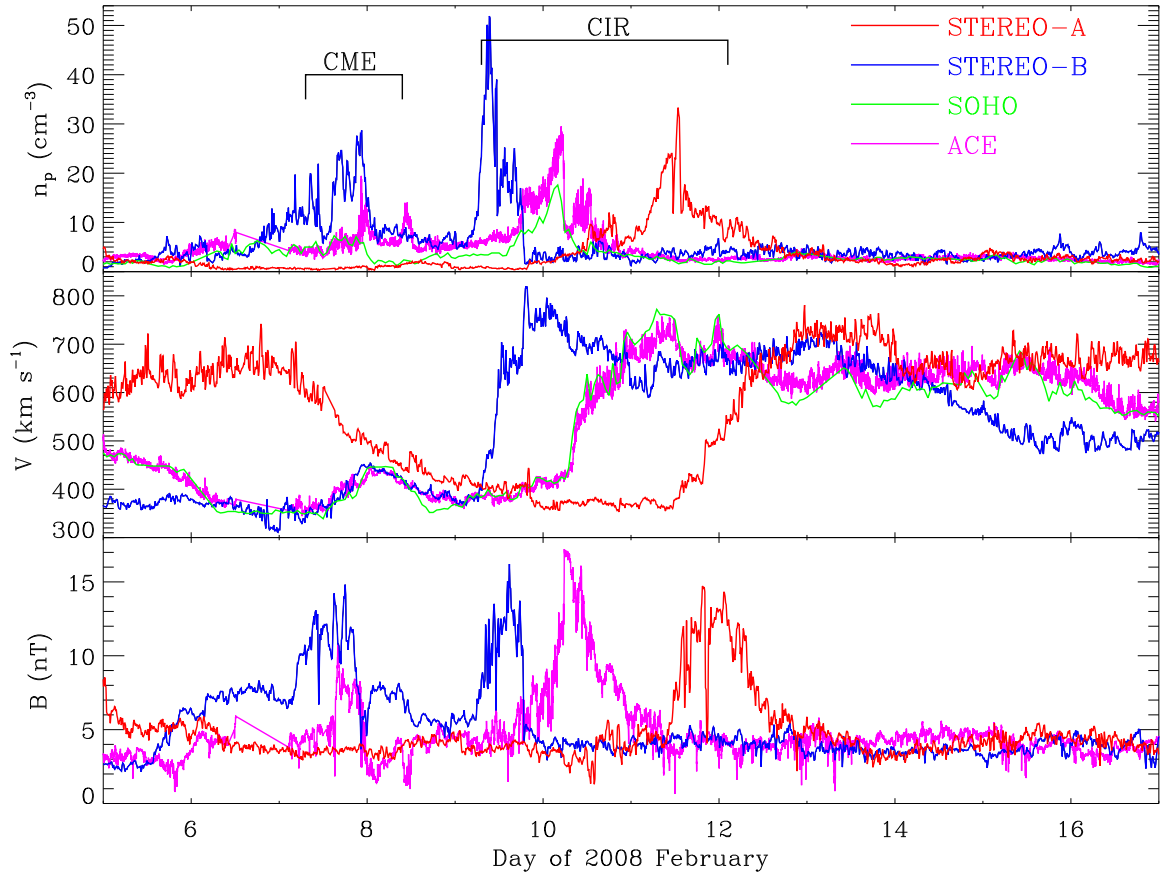}
\caption{Proton density, solar wind velocity, and magnetic field
  strength are plotted versus time using data from the PLASTIC and IMPACT
  instruments on STEREO A and B.  Also included are data from ACE and
  the CELIAS instrument on SOHO, both residing at Earth's L1 Lagrangian
  point.  The 2008 Feb.~4 CME is observed by STEREO-B on Feb.~7, and much
  more weakly by ACE and SOHO/CELIAS.  It is not seen at all by STEREO-A.
  A CIR is observed a couple days after the CME by STEREO-B, at a later
  time by SOHO/CELIAS and ACE, and later still by STEREO-A as the
  structure rotates past the various spacecraft.}
\end{figure}

\begin{figure}[p]
\plotone{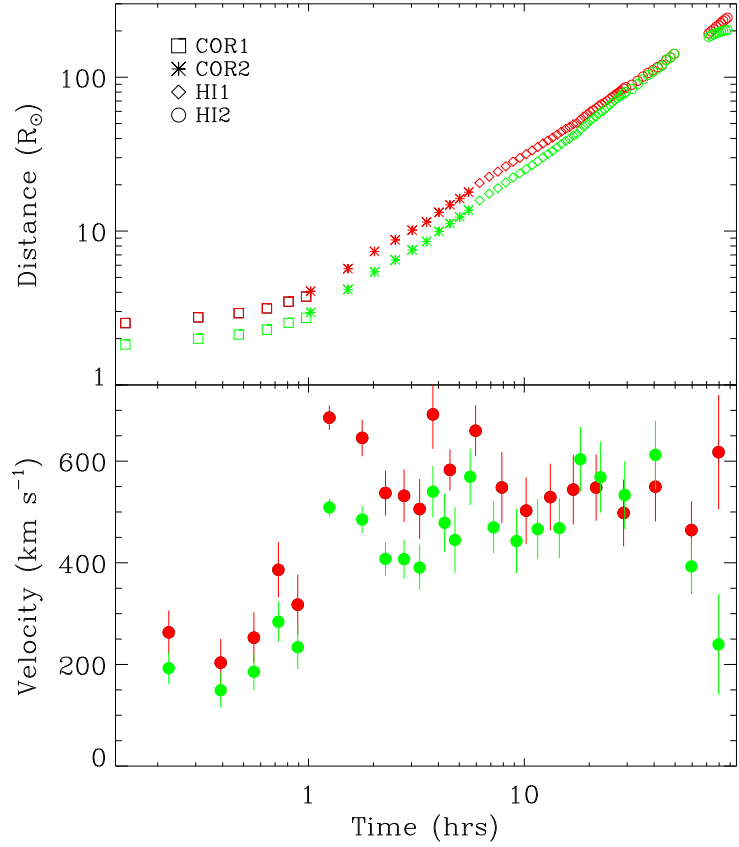}
\caption{The top panel shows two different versions of the
  distance-vs.-time plot for the leading edge of the Feb.~4 CME,
  computed using two different methods to get from measured elongation
  angle to physical distance from Sun-center.  The green measurements
  assume the ``Point-P'' method (equation 1), and the red
  data points assume the ``Fixed-$\phi$'' method (equation 2).
  The symbols indicate which SECCHI imager on STEREO-A is responsible
  for the measurement.  The bottom panel shows velocities computed
  from the distance measurements.}
\end{figure}

\begin{figure}[p]
\plotone{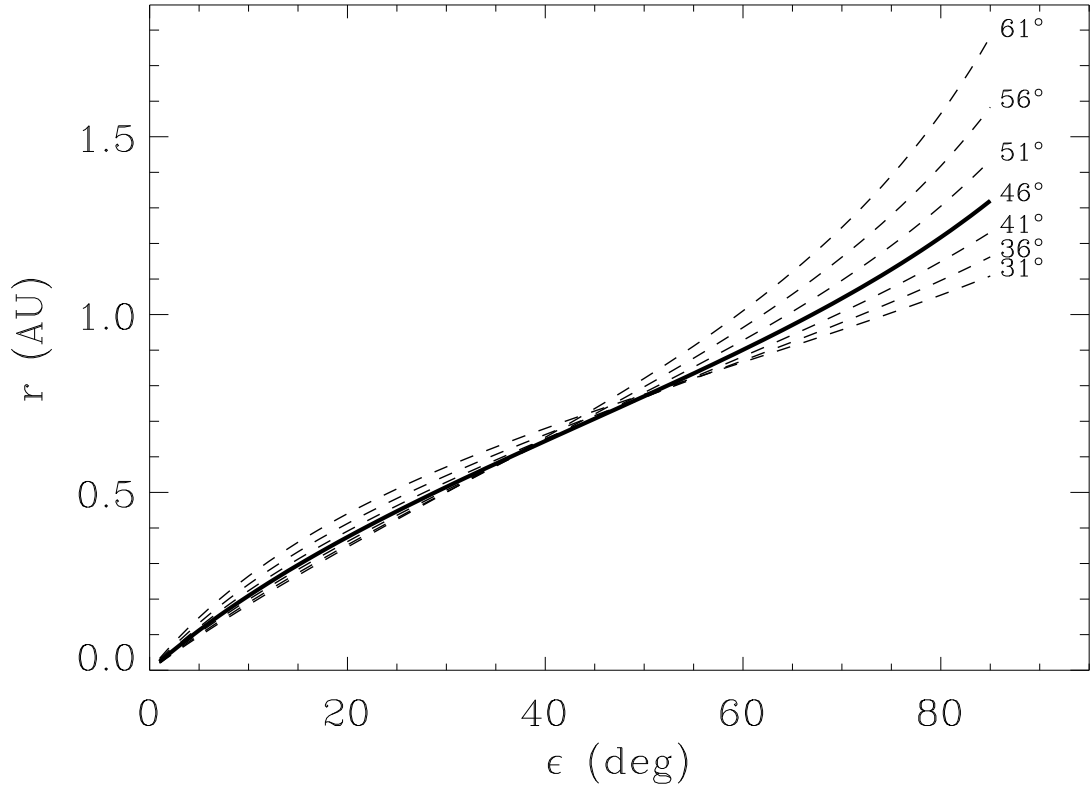}
\caption{A plot of inferred distance from Sun-center ($r$) as a function
  of measured elongation angle $\epsilon$, for seven values of the CME
  trajectory angle $\phi$, using equation (2).  The $\phi=46^{\circ}$
  curve is emphasized since that is the trajectory angle suggested
  by the flare location (see Fig.~2).}
\end{figure}

\begin{figure}[p]
\plotone{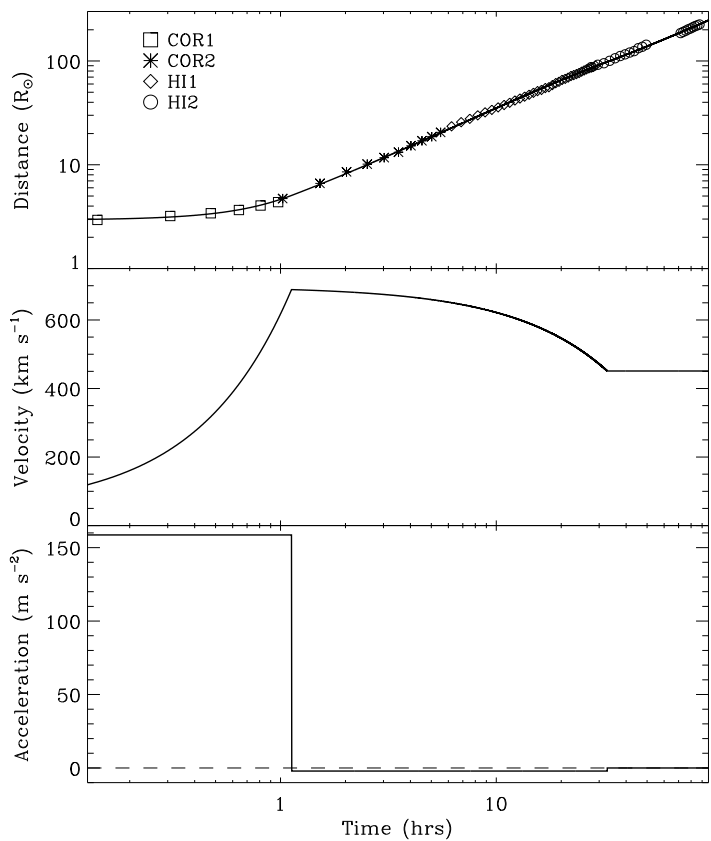}
\caption{The top panel shows the distance from Sun-center of the
  leading edge of the 2008 Feb.~4 CME as a function of time,
  assuming the CME trajectory angle is $\phi=38^{\circ}$ from the
  line of sight.  The $t=0$ time is 8:36 UT, roughly when the flare
  associated with this CME begins.  The symbols indicate which SECCHI
  imager on STEREO-A is responsible for the measurement.  The data
  points are fitted with a simple kinematic model assuming an initial
  acceleration phase, a second deceleration phase, and then a
  constant velocity phase.  The best fit is shown as a solid line
  in the top panel.  The bottom two panels show the velocity
  and acceleration profiles suggested by this fit.}
\end{figure}

\end{document}